\documentclass[preprint,tightenlines,eqsecnum,floats,aps,amsmath,amssymb,nofootinbib,prd,showpacs]{revtex4}

\usepackage{amssymb}
\usepackage{stmaryrd}
\usepackage{amsmath}
\usepackage{amsfonts}
\usepackage{mathrsfs}
\usepackage{CJK}
\usepackage{amsmath,amssymb,amsfonts}
\usepackage{graphicx}
\usepackage{subfigure}
\usepackage{color} 
\usepackage{fancyhdr}
\usepackage{hyperref} 
\newcommand{\dd}{\mathrm{d}}
\newcommand{\ee}{\mathrm{e}}
\newcommand{\ii}{\mathrm{i}}
\newcommand{\pdo}[2]{\frac{\partial #1}{\partial #2}}
\newcommand{\pd}[0]{\partial}
\newcommand{\grad}[0]{\nabla}

\def\be{\begin{equation}}
\def\eeq{\end{equation}}
\def\ba{\begin{eqnarray}}
\def\ea{\end{eqnarray}}

\usepackage{setspace}
\begin{document}
	\title{Tensor Perturbations and Thick Branes in Higher Dimensional Gauss-Bonnet Gravity}
	\date{\today}
	\author{Yongqun Xu}
	\affiliation{Department of Physics, South China University of Technology, Guangzhou 510641, China}
	\author{ Xiangdong Zhang\footnote{Corresponding author. scxdzhang@scut.edu.cn}}
	\affiliation{Department of Physics, South China University of Technology, Guangzhou 510641, China}
	
	\begin{abstract}	
		A thick brane model with Gauss-Bonnet action in a homogeneous anisotropic $D=(4+1+d)$ spacetime is studied. By choosing a concrete metric ansatz, we show that this spacetime is stable against linear tensor perturbations under certain conditions. The graviton zero modes are also given. Besides, we examine a particular example in six-dimensional spacetime with a given warp factor. The coupled background scalar field and its potential are solved analytically. Furthermore, the
effective potential of the Kaluza-Klein modes of the graviton is also discussed. We found that the effective potential can have singularities under certain conditions, which are related to the non-differentiability of the graviton zero modes.

	\end{abstract}
	\maketitle
	\section{Introduction}
	General Relativity (GR) is the most successful gravitational theory which has been proposed for over one century. Though GR has been tested by many experiments, there are still a few unknown issues such as dark energy and dark matter that remain a mystery which indicates GR is not perfect yet. These issues strongly motivate us to consider the gravity theory beyond GR. During the past several decades, many schemes of modified gravity have been put forward\cite{clifton_modified_2012}. Among all the potential directions, the extra dimension scenario is attracted increasing attention. Historically, the very first extra-dimensional gravity model is the so-called Kaluza-Klein (KK) theory which aims to unify Einstein's gravity and Maxwell's electromagnetism by introducing an extra compact circle spacial dimension. The radius scale is small enough that it can not be detected. Moreover, extra dimensions are also preferred by the string theory.
	
	On the other hand, the extra dimension could also be very large and spread infinitely. This possibility has been realized in the brane world model \cite{langlois_brane_2002,liu_introduction_2017}. An outstanding advantage of the brane world scenario is that it could nicely solve the hierarchy problem \cite{arkani-hamed_hierarchy_1998,randall_large_1999,yang_gravity_2012,barbosa-cendejas_mass_2014,guo_localization_2013,hundi_fermion_2011}. Nowadays, most of the related works on brane world model are carried out in five-dimensional spacetime. However, in five dimensions, the pure gravitational trapping mechanism of vector
fields is still problematic and there is no remarkable proposal to solve the fermion mass hierarchy
in the Standard Model yet. To overcome these mentioned issues, one may resort to higher dimensional brane world models. Particularly, brane solutions of a higher dimensional $f(R)$
gravity with a real scalar field  has been proposed recently \cite{cui_tensor_2020,zhong_tensor_2011,gu_full_2017,zhong_pure_2016,dzhunushaliev_thick_2020,PhysRevD.92.024011,cui_linear_2018,gu_stable_2018}. Note that in the $f(R)$ brane world model, one usually needs to solve higher derivative equations. One way to avoid this issue is to consider a specific higher-order polynomial combination of the Riemann tensor such that the field
equation remains two derivatives. The resulted simplest option is the famous Gauss-Bonnet gravity. The Gauss-Bonnet term is particularly suitable for higher dimensional extension of GR because it is a purely topological term in four dimensions and does not contribute to the classical equation of motion. However, this term becomes dynamic in higher dimensions. Besides, such quadratic Gauss-Bonnet terms of curvatures also appears as a 1-loop correction of heterotic
string theory \cite{GW86}. Hence inspired by these works, in this paper, we pursue to investigate the Gauss-Bonnet brane world model. Related works of Gauss-Bonnet gravity in brane world model can be also seen in the literature\cite{giovannini_thick_2001}.

Moreover, it is well known that the stability of thick branes is a very significant issue since
brane systems should be stable at least under linear perturbations. This issue has been extensively investigated \cite{cui_tensor_2020,zhong_tensor_2011,giovannini_thick_2001,giovannini_gauge-invariant_2001,giovannini_curvature_2005,zhong_pure_2016,PhysRevD.92.024011,cui_linear_2018,gu_stable_2018}. In addition, to match the gravitational experiments, the four-dimensional Newtonian potential should be recovered,
which means that the zero modes of the graviton should be localized on the brane. Hence, we also
devoted to studying the stability of tensor perturbations and the zero modes of the graviton of Gauss-Bonnet brane world model in this paper.

This paper is organized as follows: After introduction, we write down the D-dimensional Gauss-Bonnet action in the Section \ref{begin}, and then we give a particular anisotropic metric with two warp factors. The field equations under this metric ansatz are also given. In Section \ref{delta}, we mainly focus on the linear stability by computing the tensor perturbation equations. With these equations in hand, we solve the field equations analytically with a given scalar field and discuss the main property of the solution in Section \ref{6d}. Some necessary technical details can be found in the Appendix \ref{PQ}.
	
	Through out the paper we adopt the following symbol conventions: We use the lowercase Latin letters $ a,b,c,... $ running over $ 0,1,2,3,4,5... $ to stand for the indices of the $D(=4+1+d)$-dimensional spacetime. The lowercase Greek letters $ \mu,\nu,\lambda,\sigma:0,1,2,3$ represent our four-dimensional Minkowski spacetime $\mathcal{M}_4$ coordinates. Specially we use $ x^4=y $ as the fifth extra spacial coordinate $\mathcal{R}_1$. The capital Latin $ A,B,C...: 5,6,7... $ live in the higher extra dimensional Euclidean space $\mathcal{E}_d$ with $d=1,2,...$. Moreover, through out the paper, we use the definition $ R_{a b c}{}^{d}=\pd_b\Gamma^{d}{}_{a c}-\pd_a\Gamma^{d}{}_{b c}+\Gamma^{e}{}_{c a} \Gamma^{d}{}_{b e}-\Gamma^{e}{}_{c b} \Gamma^{d}{}_{a e} $ and the spacetime signature $ (-,+,+...) $.

	\section{Action, Field equation And Metric}\label{begin}
	Our discussion in this article is based on the following D-dimensional Einstein-Hilbert action with a Gauss-Bonnet term $ \mathcal{L}_{GB}=R^2-4R_{ab}R^{ab}+R_{abcd}R^{abcd}$ as
\begin{eqnarray}\label{action}
S=\frac{1}{2\kappa_D^2}\int d^D x \sqrt{-g}\left(R+\alpha\mathcal{L}_{GB}\right)+S_m,
\end{eqnarray}
where $g=\det g_{ab}$, $\kappa^2_D=8\pi G^{(D)}$ with the $ G^{(D)} $ being the gravitional constant in D-dimensional spacetime, and $\alpha$ is the Guass-Bonnet coupling constant which has dimension of $[lengh]^2 $. By using this action, we can directly derive the field equations by varing the action with respect to the metric tensor $g_{ab}$:
	\begin{eqnarray}\label{eom}
		G_{ab}+\alpha H_{ab}=\kappa_D^2 T_{ab},
	\end{eqnarray} where $ T_{ab}=-\dfrac{2}{\sqrt{-g}} \dfrac{\partial\left(\sqrt{-g} \mathcal{L}_{m}\right)}{\partial g^{ab}} $ is the energy-momentum tensor, and the lovelock tensor $ H_{ab} $ is coming from the contribution of the Gauss-Bonnet term \cite{clifton_modified_2012}:
	\begin{eqnarray}\label{Hab}
		H_{ab}=-4 R_{a}{}^{c} R_{bc} + 2 R_{ab} R - 4 R^{cd} R_{acbd} + 2 R_{a}{}^{cde} R_{bcde}-\frac12g_{ab}\mathcal{L}_{GB}.
	\end{eqnarray}
	
	Parallel to the previous study such as in $ f(R) $ brane world model \cite{cui_tensor_2020}. In this paper, we also consider a  special anisotropic $ (D=4+1+d) $-dimensional bulk manifold $\mathcal{M}=\mathcal{M}_4\times\mathcal{R}_1\times\mathcal{E}_d $, which is constructed by three subspaces:
	
	\begin{itemize}
		\item A 4-dimentional Minkowski spacetime $\mathcal{M}_4$,
		\item An extra 1-dimensional noncompact space  $\mathcal{R}_1$,
		\item An extra d-dimensional Euclidean space $ \mathcal{E}_d $.
	\end{itemize} Naturally, there should exist curved brane (like $ \dd S $ and $ A\dd S $ brane) as well as flat brane. In this paper, we are particularly interested in a flat and static brane, of which the line element can be written as:
	\begin{eqnarray}\label{metric}
		\dd s^2=a^2(y)\eta_{\mu\nu}\dd x^\mu\dd x^\nu+\dd y^2+b^2(y)\delta_{AB}\dd x^A\dd x^B
	\end{eqnarray}
	where $ a(y)=\ee ^{A(y)}$ and $b(y)=\ee^{B(y)} $ are two warped factors. The Gauss-Bonnet term in this metric can be computed directly as

		\begin{eqnarray}\label{gbtofmetric}
			\mathcal{L}_{GB}=&&\frac{24 a'^4}{a^4}+\frac{96 a'^2 a''}{a^3}+d \left(\frac{48 a'^2 b''}{a^2 b}+\frac{96 a'^3 b'}{a^3
				b}+\frac{96 a' a'' b'}{a^2 b}+(d-1) \left(\frac{16 a'' b'^2}{a b^2}
			\right.\right.\notag\\&& \left.\left.
			+\frac{72a'^2 b'^2}{a^2 b^2}+\frac{32 a' b' b''}{ab^2}+(d-2) \left(\frac{16 a' b'^3}{a b^3}+\frac{4 b'^2 b''}{b^3}+(d-3)\frac{
				b'^4}{b^4}\right)\right)\right)
		\end{eqnarray}
	Some higher order terms will arise as $ d $ grows larger till $ d=4 $. Moreover, the field equations under this metric ansatz are:
	\begin{small}
	\begin{eqnarray}\label{eomuv}	
			&&\frac{3 a''}{a}+\frac{3 a'^2}{a^2}+d \left(\frac{3 a' b'}{a b}+\frac{b''}{b}-\frac{b'^2}{2
				b^2}\right)+\frac{d^2 b'^2}{2 b^2}
			-\alpha \left[
			\frac{12 a'^2a''}{a^3}
			-d \left(\frac{6 a'' b'^2}{ab^2}-\frac{12 a'^2 b''}{a^2 b}-\frac{12 a' b'^3}{a b^3}
			\right.\right.\notag\\&&\left.
			+\frac{18 a'^2 b'^2}{a^2b^2}-\frac{12 a'^3 b'}{a^3 b}+\frac{12 a' b' b''}{a b^2}-\frac{24 a' a''b'}{a^2 b}+\frac{3 b'^4}{b^4}-\frac{4 b'^2 b''}{b^3}\right)
			-d^2 \left(\frac{6 b'^2 b''}{b^3}-\frac{18 a'^2 b'^2}{a^2 b^2}-\frac{11b'^4}{2 b^4}
			\right.\notag\\
			&&\left.\left.
			+\frac{18a' b'^3}{a b^3}-\frac{6 a'' b'^2}{a b^2}-\frac{12 a' b' b''}{ab^2}\right)				
			+d^3 \left(\frac{6 a' b'^3}{a b^3}-\frac{3b'^4}{b^4}+\frac{2 b'^2 b''}{b^3}\right)
			+\frac{d^4 b'^4}{2 b^4}
			\right]
			=\frac{\kappa_D^2\eta^{\mu\nu}T_{\mu\nu}}{4a^2}
	\end{eqnarray}
	\begin{eqnarray}\label{eomyy}	
			&&\frac{6 a'^2}{a^2}+d \left(\frac{4 a'b'}{a b}-\frac{b'^2}{2 b^2}\right)+d^2\frac{ b'^2}{2 b^2}+
			\alpha\left[
			-\frac{12 a'^4}{a^4}-\frac{d^4 b'^4}{2 b^4}
			+d \left(-\frac{16a' b'^3}{a b^3}+\frac{36 a'^2 b'^2}{a^2 b^2}-\frac{48 a'^3 b'}{a^3 b}
						\right.\right.\notag\\&&\left.
			+\frac{3b'^4}{b^4}\right)
			+d^2 \left(\frac{24 a'b'^3}{a b^3}-\frac{36 a'^2 b'^2}{a^2 b^2}-\frac{11 b'^4}{2 b^4}\right)
			\left.+d^3 \left(\frac{3 b'^4}{b^4}-\frac{8 a' b'^3}{a b^3}\right)
			\right]={\kappa_D^2} {T_{yy}}
	\end{eqnarray}
	\begin{eqnarray}\label{eomAB}	
			&&\frac{b'^2}{b^2}-\frac{b''}{b}+\frac{6 a'^2}{a^2}+
			\frac{4 a''}{a}-\frac{4 a' b'}{a b}+d \left(\frac{4 a' b'}{a
				b}+\frac{b''}{b}-\frac{3 b'^2}{2 b^2}\right)+d^2\frac{ b'^2}{2 b^2}+
			\alpha  \left[\frac{24 a'^2 b''}{a^2 b}-\frac{16 a''b'^2}{a b^2}
			\right.\notag\\&&
			+\frac{48 a'^3 b'}{a^3 b}-\frac{72 a'^2b'^2}{a^2 b^2}+\frac{48 a' b'^3}{a b^3}-\frac{32 a' b' b''}{a b^2}-\frac{12a'^4}{a^4}+\frac{48 a' a'' b'}{a^2 b}
			+d \left(\frac{24 a'' b'^2}{ab^2}-\frac{24 a'^2 b''}{a^2 b}-\frac{88 a' b'^3}{a b^3}
			\right.\notag\\&&\left.
			+\frac{108 a'^2b'^2}{a^2 b^2}-\frac{48 a'^3 b'}{a^3 b}+\frac{48 a' b' b''}{a b^2}-\frac{48a' a'' b'}{a^2 b}+\frac{23 b'^4}{b^4}-\frac{22 b'^2 b''}{b^3}\right)
			\notag\\&&
			+d^2 \left(-\frac{8 a'' b'^2}{a b^2}
			+\frac{48a' b'^3}{a b^3}-\frac{36 a'^2 b'^2}{a^2 b^2}-\frac{16 a' b' b''}{a
				b^2}-\frac{35 b'^4}{2 b^4}+\frac{12 b'^2 b''}{b^3}\right)
		\notag\\&&\left.
			-\frac{10 b'^4}{b^4}+\frac{12 b'^2b''}{b^3}-\frac{48 a'^2a''}{a^3}-\frac{d^4 b'^4}{2 b^4}+d^3 \left(-\frac{8 a' b'^3}{a b^3}
			+\frac{5b'^4}{b^4}-\frac{2 b'^2 b''}{b^3}\right)
			\right]=\frac{\kappa_D^2\eta^{AB}}{db^2} {T_{AB}},
	\end{eqnarray}	\end{small} where the prime $'$ denotes the derivative with respect to the extra-dimensional coordinate $ y $. With a given matter distribution we can solve the equations above. But before we proceed, we should wonder whether this set of equations gives a reasonable solution, or under what condition they do so. To this aim, we need to first analyze the behavior under linear perturbation carefully.

	\section{Linear Stability Under Tensor Perturbation}\label{delta}
	The perturbed equation can be very complicated for such a Gauss-Bonnet action\eqref{action} and bulk metric \eqref{metric}. The perturbed tensor equations of motion can be formally written as:
	\begin{eqnarray}\label{abspteq}
		\delta G_{ab}+\alpha \delta H_{ab}=\kappa_D^2 \delta T_{ab}.
	\end{eqnarray} Usually, a linear perturbation can be decomposed into scalar, vector and tensor parts. Since these three perturbations are decoupled to each other. Hence here for simplicity, we just let alone the others and just focus on the tensor perturbation. Note that the tensor perturbation can impose various kinds of gauge conditions. Here we consider the most popular one which is the so-called the transverse-traceless(TT) gauge condition:
	\begin{eqnarray}\label{h}
		\dd s^2=a^2(y)(\eta_{\mu\nu}+h_{\mu\nu})\dd x^\mu\dd x^\nu+\dd y^2+b^2(y)\delta_{ij}\dd x^i\dd x^j,
	\end{eqnarray}
	where $ h_{\mu\nu} $ obeys the TT gauge condition:
	\begin{eqnarray}\label{gauge}
		\pd_\mu h_\nu^\mu,\quad  h=\eta^{\mu\nu}h_{\mu\nu}=0.
	\end{eqnarray} The tensor perturbation of $ f(R) $ brane-world model with TT gauge is carried out in \cite{cui_tensor_2020,zhong_tensor_2011}. Similarly we compute the perturbed quantities step by step, from the perturbed Christoffel connection to the Riemann Curvature. The non-vanishing perturbed Christoffel connection are listed below:
	\begin{eqnarray}\label{dchris}
			\delta \Gamma^{\lambda}{}_{\mu\nu}&=&\frac{1}{2}(\pd_\mu h_{\nu}^{\lambda}+\pd_\nu h_{\mu}^{\lambda}-\pd^{\lambda}h_{\mu\nu})\notag\\
			\delta \Gamma^{y}{}_{\mu\nu}&=&-aa'h_{\mu\nu}-\frac{a^2}{2}\pd_yh_{\mu\nu}\notag\\
			\delta \Gamma^{A}{}_{\mu\nu}&=&\frac{1}{2}\frac{a^2}{b^2}\pd^{A}h_{\mu\nu}\notag\\
			\delta \Gamma^{\lambda}{}_{\mu y}&=&\frac{1}{2}\pd_yh_{\mu}^{\lambda}\notag\\
			\delta \Gamma^{\lambda}{}_{\mu A}&=&\frac{1}{2}\pd_Ah_{\mu}^{\lambda}
	\end{eqnarray} By directly using the formula $ \delta {R_{abc}}^d=\nabla_b\delta{\Gamma^d}_{ac}-\nabla_a\delta{\Gamma^d}_{bc} $	, one can computed all of the needed perturbation quantities which can be found in Appendix \ref{PQ}. For example, the pertured Ricci tensor can be written as :
	\begin{eqnarray}\label{dRuv}
			\delta R_{\mu \nu}=-\frac{a^{2}}{2}
			\left[
			\left( \frac{4a'}{a}+ \frac{db'}{b}\right)\pd_y
			+\left(\pd_y^2+\frac{\square^{(4)}}{a^{2}}  +\frac{\hat{\Delta}^{(d)}}{b^{2}}  \right)
			+2\left( \frac{3a'^{2}}{a^{2}}+\frac{a''}{a}+\frac{da'b'}{ab}\right) \right]h_{\mu \nu},
	\end{eqnarray} where $ \square^{(4)}=\eta^{\mu\nu}\pd_\mu\pd_\nu,\quad  \hat{\Delta}^{(d)}=\delta^{ij}\pd_i\pd_j $ are the d'Alembert and Laplace operator of Minkowski space $\mathcal{M}_4$ and extra d-dimensional spacial Euclidean space $\mathcal{E}_d$ respectively. Finally,  after tedious but straightforward computation, we end up with $ (\mu,\nu) $ components of the perturbed field equations:
\begin{small}\begin{eqnarray} \label{2ndpert}
	&&\left[-\frac{1}{2 a^2}+\alpha  \left(\frac{2 a''}{a^3}+d \left(\frac{2 a' b'}{a^3 b}+\frac{2 b''}{a^2 b}-\frac{b'^2}{a^2 b^2}\right)+\frac{d^2 b'^2}{a^2 b^2}\right) \right]\square^{(4)}h_{\mu\nu}\notag\\
	&&-\left[\frac{1}{2 b^2}-\alpha  \left(\frac{4 a''}{a b^2}-\frac{4 a' b'}{a b^3}+\frac{2 a'^2}{a^2 b^2}-\frac{2 b''}{b^3}+\frac{2 b'^2}{b^4}+d \left(\frac{4 a' b'}{a b^3}+\frac{2 b''}{b^3}-\frac{3 b'^2}{b^4}\right)+d^2\frac{ b'^2}{b^4}\right)\right]\hat{\Delta}^{(d)}h_{\mu\nu}\notag\\
	&&+\left[ -\frac{2 a'}{a}-\frac{d b'}{2 b}
	+\alpha \left(
	\frac{4 a' a''}{a^2}+\frac{4 a'^3}{a^3}
	+d \left(\frac{4 a'' b'}{a b}+\frac{4 a' b''}{a b}-\frac{8 a' b'^2}{a b^2}+\frac{14 a'^2 b'}{a^2 b}+\frac{2 b'^3}{b^3}-\frac{2 b' b''}{b^2}\right)
	\right.\right.\notag\\&&\left.\left.
	+d^2 \left(\frac{8 a' b'^2}{a b^2}-\frac{3 b'^3}{b^3}+\frac{2 b' b''}{b^2}\right)
	+\frac{d^3 b'^3}{b^3}
	\right)\right]h'_{\mu\nu}\notag\\
	&&+\left[-\frac{1}{2}+ \alpha  \left(\frac{2 a'^2}{a^2}+d \left(\frac{4 a' b'}{a b}-\frac{b'^2}{b^2}\right)+\frac{d^2 b'^2}{b^2}\right)\right]h''_{\mu\nu}=0.
\end{eqnarray}\end{small}
	We need to add that, the $ h_{\mu\nu} $ term of Eq.\eqref{2ndpert} has been eliminated with the consideration of the field equation\eqref{eomuv}. We will apply some mathematical methods to get some physical results from the equation above. First of all, we introduce a coordinate transformation $ \dd y=a\dd z,\ \dd x^A=\dfrac a b\dd w^A $ to obtain a conformal-flat metric:
	\begin{eqnarray}\label{conflat}
		\dd s^2=a^2(z)(\eta_{\mu\nu}\dd x^\mu\dd x^\nu+\dd z^2+\delta_{AB}\dd x^A\dd x^B).
	\end{eqnarray} Then the perturbation equation \eqref{2ndpert}  in such conformal flat coordinate reads:
	\begin{eqnarray} \label{2ndpertnew}
			&& \left[\frac{1}{2 a^2}-\alpha  \left(-\frac{2 a'^2}{a^6}+\frac{2 a''}{a^5}+d \left(-\frac{b'^2}{a^4 b^2}+\frac{2 b''}{a^4
				b}\right)+d^2\frac{ b'^2}{a^4 b^2}\right)\right]\square^{(4)}h_{\mu\nu}+\left[\frac{1}{2 a^2}-\alpha  \left(\frac{4a''}{a^5}
			\right.\right.\notag\\&&\left.\left.
			-\frac{2 a'^2}{a^6}-\frac{2 a' b'}{a^5 b}+\frac{2 b'^2}{a^4 b^2}-\frac{2 b''}{a^4 b}+d \left(\frac{2 a' b'}{a^5 b}-\frac{3 b'^2}{a^4 b^2}+\frac{2 b''}{a^4
				b}\right)+\frac{d^2 b'^2}{a^4 b^2}\right)\right]\hat{\Delta}^{(d)}h_{\mu\nu}+\left[\frac{3 a'}{2 a^3}
			\notag\right.\\&&\left.
			+\frac{d b'}{2 a^2 b}-\alpha  \left(-\frac{2 a'^3}{a^7}+\frac{4 a' a''}{a^6}+d \left(\frac{2
				a'^2 b'}{a^6 b}-\frac{5 a' b'^2}{a^5 b^2}+\frac{2 b'^3}{a^4 b^3}+\frac{4 b' a''}{a^5 b}+\frac{4
				a' b''}{a^5 b}-\frac{2 b' b''}{a^4 b^2}\right)
			\right.\right.\notag\\&&\left.\left.
			+d^2 \left(\frac{5 a' b'^2}{a^5 b^2}-\frac{3 b'^3}{a^4b^3}+\frac{2 b' b''}{a^4 b^2}\right)+\frac{d^3 b'^3}{a^4 b^3}\right)\right]h'_{\mu\nu}
			\notag\\&&
			+\left[\frac{1}{2 a^2}-\alpha  \left(\frac{2 a'^2}{a^6}+d \left(\frac{4 a' b'}{a^5 b}-\frac{b'^2}{a^4
				b^2}\right)+\frac{d^2 b'^2}{a^4 b^2}\right)\right]h''_{\mu\nu}=0.
	\end{eqnarray} This is a second-order linearized tensor equation that can be abstractly written as follows:
	\begin{eqnarray}\label{abshuv}
		J(z)\square^{(4)}h_{\mu\nu}+K(z)\hat{\Delta}^{(d)}h_{\mu\nu}+2P(z)\pd_zh_{\mu\nu}+Q(z)\pd_z^2h_{\mu\nu}=0
	\end{eqnarray}
	where $ J,K,P,Q $ are functions of coordinate $ z $ defined by Eq.\eqref{2ndpertnew}. Naturally, are we want to solve the equation, we are willing to decompose it into three different parts respectively:
	\begin{eqnarray}
		h_{\mu\nu}=\epsilon_{\mu\nu}(x^\mu)\Phi(z)\xi(x_A).
	\end{eqnarray} We can firstly obtain the $ (\mu,\nu) $ part, which is a Klein-Gordon equation:
	\begin{eqnarray}\label{KG}
		(\square^{(4)}-m^2)\epsilon_{\mu\nu}(x^\mu)=0
	\end{eqnarray}	
	and the Helmholtz equation for the extra d-dimentions:
	\begin{eqnarray}\label{Helmholtz}
		\left(\hat{\Delta}^{(d)}+l^2\right)\xi(x_A)=0.
	\end{eqnarray}
	What we left behind is a second-order ordinary differential equations for the extra fifth dimension:
	\begin{eqnarray}\label{yy}
		\left(2\frac{P}{Q}\pd_z+\pd_z^2\right)\Phi(z)=\left(\frac{K}{Q}l^2-\frac{J}{Q}m^2\right)\Phi(z).
	\end{eqnarray} We can always eliminate the one-order derivative $\pd_z$ term with one more factor $\Phi(z)=\psi(z)\exp(-\int P/Q\dd z)$, one can check that the Eq.\eqref{yy} will be turned into a Shrodinger-like equation\cite{gu_full_2017}:
	\begin{eqnarray}\label{schrodinger}
		\left(-\pd_z^2+U(z) \right)\psi(z)=\left(\frac{J}{Q}m^2-\frac{K}{Q}l^2\right)\psi(z)
	\end{eqnarray}
	where $ U(z) $ is the effective potential defined as
	\begin{eqnarray}\label{ep}
		U(z)=\left(\frac{P}{Q}\right)^2+\pd_z\left(\frac{P}{Q}\right).
	\end{eqnarray}
	After that, we we can introduce a superpential $ W(x)=-P/Q $ and re-written the Hamiltonian of Eq.\eqref{schrodinger} into two operators $ A^\dagger=\pd_z+W$ and $A=-\pd_z+W $ as what we do in supersymmetric quantum mechanics \cite{cooper_supersymmetry_1995}:
	
	\begin{eqnarray}
		A^\dagger A\psi(z)=\left(\frac{J}{Q}m^2-\frac{K}{Q}l^2\right)\psi(z).
	\end{eqnarray} With all this process, we can finally give the condition that makes sure the stability under linear perturbation. It immediately follows that we must have $ m^2\ge0 $ and $ l^2\ge0 $ in Eq.\eqref{KG} and Eq.\eqref{Helmholtz} for stability. And more importantly, we must have the graviton zero mode normalizable to avoid tachyon states.
	
	Setting $ m^2=l^2=0 $ in Eq.\eqref{schrodinger}, or $ A\psi(z)=0 $, at least, we can obtain the ground state, or the graviton zero modes of this system:
	\begin{eqnarray}\label{key}
		\psi_0(z)\propto \exp\int\frac{P}{Q}\dd z.
	\end{eqnarray}	With the help of these equations, we can set several restrictions for the perturbation. In the following, we will show the whole procedure through a specific example in the next section.

	\section{Brane Solution In 6d Spacetime}\label{6d}
	
	After perturbation analysis, we now turn to search for a particular brane solution. We choose $ d=1 $ as a non-trivial example. In 6-dimensions or the case $ d=1 $, the line element reads
	\begin{eqnarray}\label{metric6d}
		\dd s^2=a^2(y)\eta_{\mu\nu}\dd x^\mu\dd x^\nu+\dd y^2+L^2b^2(y)\dd \theta^2.
	\end{eqnarray} where $\theta\in[0,2\pi)$ is the coordinate of the sixth dimension which is compact.
	
\begin{figure}[htbp]\centering
		\includegraphics[scale=0.8]{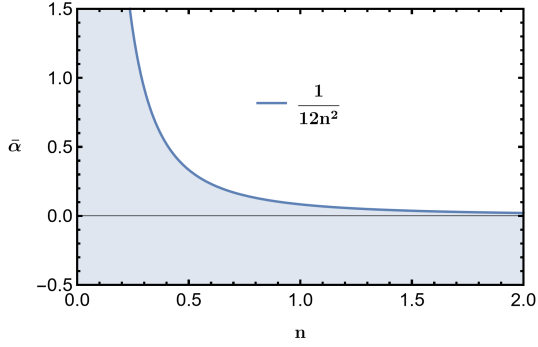}
		\caption{The real-value condition for the background scalar field sets a restriction for the Gauss-Bonnet coupling constant $ \alpha $. Unlike $ f(R) $ brane, the parameter $ \bar{\alpha} $ in this context can be sufficiently large for small $ n $. }\label{domainofa}
	\end{figure}

	Consider a background scalar field $ \phi(y) $ which is only the function of $ y $ for a static flat brane, with the following action of the matter:
	\begin{eqnarray}\label{sm}
		S_m=\int\dd^6x\sqrt{-g}\left(-\frac12\pd^a\phi\pd_a\phi-V(\phi)\right),
	\end{eqnarray} where $ V(\phi) $ is the self-interacting potential. Set $ d=1 $ in Eq. \eqref{eomuv},\eqref{eomyy},\eqref{eomAB}, the components of the field equations can be written  as follows :
	\begin{eqnarray}
		\label{eom6d}
		&&\frac{3 a' b'}{a b}+\frac{3 a'^2}{a^2}+\frac{3
			a''}{a}+\frac{b''}{b}-12\alpha  \left(\frac{ a'^2 b''}{a^2 b}+\frac{2 a' a'' b'}{a^2 b}+\frac{ a'^3 b'}{a^3
			b}+\frac{ a'^2 a''}{a^3}\right) =-\kappa_6^2 \left[V(\phi)+\frac12\phi'^2\right]\notag\\
		&&\frac{4 a' b'}{a b}+\frac{6a'^2}{a^2}-12\alpha  \left(\frac{4 a'^3 b'}{a^3 b}+\frac{a'^4}{a^4}\right)= \kappa_6^2 \left[\frac12\phi '^2- V(\phi)\right]\notag\\
		&&\frac{4 a''}{a}+\frac{6a'^2}{a^2}-12\alpha  \left(\frac{4 a'^2 a''}{a^3}+\frac{a'^4}{a^4}\right)= -\kappa_6^2 \left[\frac12\phi	
		'^2+V(\phi )\right]
	\end{eqnarray}	Moreover, the equation of motion for the background scalar field is:
\begin{eqnarray}\label{bgs}
		\phi''+\left(4\frac{a'}{a}+\frac{b'}{b}\right)\phi'-\pdo{V}{\phi}=0.
	\end{eqnarray} In order to solve these equations, we give linearly combinations from Eq. \eqref{eom6d}
	\begin{eqnarray}\label{eq6d}
		&&\alpha  \left(\frac{48 a'^2 a''}{a^3}-\frac{48 a'^3 b'}{a^3 b}\right)+\frac{4 a' b'}{a
			b}-\frac{4 a''}{a} = \kappa_6^2 \phi '^2 \notag\\
		&&\alpha  \left(\frac{12 a'^2 b''}{a^2 b}+\frac{24 a' a'' b'}{a^2 b}-\frac{36 a'^3 b'}{a^3
			b}+\frac{12 a'^2 a''}{a^3}-\frac{12 a'^4}{a^4}\right)+\frac{a' b'}{a b}+\frac{3
			a'^2}{a^2}-\frac{3 a''}{a}-\frac{b''}{b} = \kappa_6^2 \phi'^2 \notag\\
		&&\alpha  \left(\frac{12 a'^2 b''}{a^2 b}+\frac{24 a' a'' b'}{a^2 b}+\frac{12 a'^3 b'}{a^3
			b}-\frac{36 a'^2 a''}{a^3}-\frac{12 a'^4}{a^4}\right)-\frac{3 a' b'}{a b}+\frac{3
			a'^2}{a^2}+\frac{a''}{a}+\frac{b''}{b} = 0.
	\end{eqnarray} Then the main perturbation equation for this case reads:
	\begin{eqnarray}
		&&\hat{\Delta}^{(d)}h_{\mu\nu} \left(\frac{1}{2 L^2b^2}-\alpha  \left(\frac{2 a'^2}{a^2 b^2}+\frac{4 a''}{a
			b^2}\right)\right)
		+\square^{(d)}h_{\mu\nu} \left(\frac{1}{2 a^2}-\alpha  \left(\frac{2 a' b'}{a^3 b}+\frac{2
			a''}{a^3}+\frac{2 b''}{a^2 b}\right)\right)
		\notag\\&&
		+h_{\mu\nu}' \left(\frac{2 a'}{a}+\frac{b'}{2 b}-\alpha
		\left(\frac{4 a'^3}{a^3}+\frac{14 a'^2 b'}{a^2 b}+\frac{4 a' a''}{a^2}+\frac{4 b'
			a''}{a b}+\frac{4 a' b''}{a b}\right)\right)\notag\\&&
		+h_{\mu\nu}''\left(\frac{1}{2}-\alpha  \left(\frac{2
			a'^2}{a^2}+\frac{4 a' b'}{a b}\right)\right) =0.
	\end{eqnarray} We choose a particular smooth function for $ a(y) $ and $ b(y) $ which is also widely adopted in the literature \cite{liu_introduction_2017,gu_full_2017,yu_gravitational_2016,cui_tensor_2020,zhong_tensor_2011,guo_tensor_2016,yang_thick_2012,csaki_universal_2000,liu_domain_2011,zhong_pure_2016,PhysRevD.92.024011,cui_linear_2018} to solve and analyze the brane. Here we assume that
	\begin{eqnarray}\label{ab}
		a(y)=b(y)=\mathrm{sech}^n(ky),
	\end{eqnarray} with $ n>0 $, so that the metric\eqref{metric} can be reduced to an $ A\dd S $ at infinity. While the parameter $ k >0$ was set to adjust the thickness of the brane. Substitute Eq.\eqref{ab} into Eq.\eqref{eq6d}, we will get the only one admissable equation as:
	\begin{eqnarray}\label{ode6d}
		\kappa_6^2 \phi'^2 =48 \alpha  k^4 n^3 \text{sech}^4(k y)+4 k^2 n \left(1-12 \alpha  k^2 n^2\right) \text{sech}^2(k y)
	\end{eqnarray} Note that it is easy to check that the ordinary differential equation :
	\begin{eqnarray}
		\kappa  f'(y)=\pm	\sqrt{a_1 (1-a_2) \text{sech}^4(k y)+a_1 a_2 \text{sech}^2(k y)}
	\end{eqnarray} has the solution:
	\begin{eqnarray}\label{math}
		f(y)=\pm \frac{\sqrt{a_1}}{\kappa  k}
		\left(
		i \left[E(i k y|a_2)-F(i ky|a_2)\right]
		+ \tanh (k y) \sqrt{a_2 \sinh ^2(k y)+1}
		\right),
	\end{eqnarray} where $ F$ and $ E $ are the first and second elliptic integrals respectively. Compare the above equation with the Eq. \eqref{ode6d} we have
	\begin{eqnarray}%
		a_1 &&=4 k^2 n, \quad a_2= 1 - 12 k^2 n^2 \alpha\\
\phi(y)&&=\pm2 \frac{\sqrt{n}}{\kappa_6} \left(\ii [E(i k y|p)-F(i k y|p )]+\tanh (k y) \sqrt{a_2 \sinh ^2(k y)+1}\right).
	\end{eqnarray} Moreover, with the help of Eq.\eqref{bgs} to obtain the potential of the scalar field as:
	\begin{eqnarray}\label{phiv}
		V(\phi(y))=2\frac{k^2n}{\kappa_6^2}[ (5 n+1) a_2 \mathrm{sech}^2(k y) +6 \alpha  k^2 n^2 (5 n+2) \mathrm{sech}^4(k y)].
	\end{eqnarray} Since the background scalar field $ \phi(y) $ should be real-valued, it proposes a restriction for the parameters in the ellipse integral, which means $ a_1>0, \ a_2\ge0 $ or :
	\begin{eqnarray}\label{alphacondi}
		\frac{1}{12n^2}\ge k^2\alpha
	\end{eqnarray} The feasible region of the Gauss-Bonnet coupling constant $ \alpha $ is shown in Fig.\ref{domainofa}
	\begin{figure}\centering
		\begin{minipage}[t]{0.48\textwidth}\centering
			\includegraphics[scale=0.8]{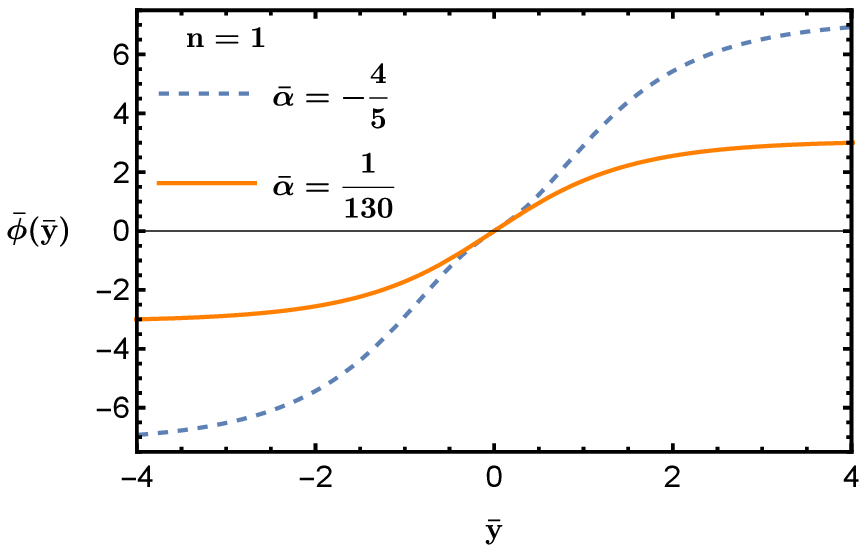}
			\caption{The background scalar field $ \bar{\phi}(\bar{y}) $.}\label{phi}
		\end{minipage}
		\begin{minipage}[t]{0.48\textwidth}\centering
			\includegraphics[scale=0.8]{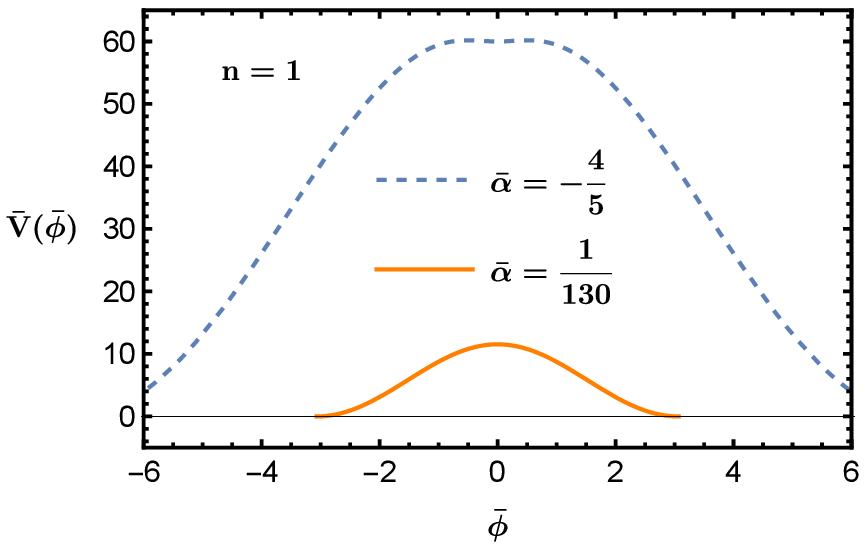}
			\caption{The potential $ \bar{V}(\bar{\phi}) $.}\label{Vphi}
		\end{minipage}
	\end{figure}
	To make things clear, we need to introduce the following dimensionless quantities:
	\begin{eqnarray}
		\bar{y}=k y,\qquad\bar{\alpha}=k^2\alpha,\qquad \bar{\phi}=\kappa_6\phi,\qquad\bar{V}=\kappa_6^2V/k^2.
	\end{eqnarray} Furthermore, the scalar field $ \phi(y) $ and its potential $ V(\phi) $ are shown in Fig.\ref{phi} and Fig.\ref{Vphi} respectively. The parameter $ n $  is set to 1 in these two figures.
	
	From Fig. \ref{phi}, we should notice that the scalar field $ \bar{\phi}(\bar{y}) $ has two kinks for small $ \bar{\alpha} $ and only one kink for large $ \bar{\alpha} $. So there must be a critical value of $ \bar{\alpha} $ which $ \bar{\phi}(\bar{y}) $ will have one more kink once $ \alpha $ exceed.
	
\begin{figure}[htbp]\centering
		\begin{minipage}[t]{0.48\textwidth}\centering
			\includegraphics[scale=0.8]{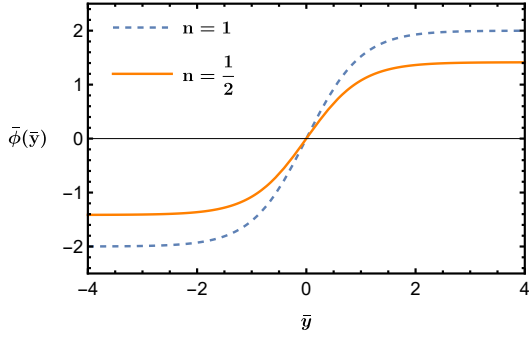}
			\caption{The scalar field $ \bar{\phi}(\bar{y}) $ when $ \bar{\alpha}=1/12n^2 $.}\label{phic}
		\end{minipage}
		\begin{minipage}[t]{0.48\textwidth}\centering
			\includegraphics[scale=0.8]{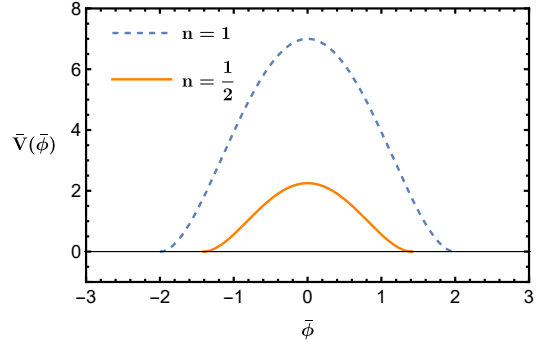}
			\caption{The potential $ \bar{V}(\bar{\phi}) $when $ \bar{\alpha}=1/12n^2$.}\label{Vphic}
		\end{minipage}
	\end{figure}

	This value can be found by solving the following equations:
	\begin{eqnarray}\label{ac}
		\left.\frac{\dd^3 \bar{\phi}}{\dd \bar{y}^3}\right|_{\bar{y}=0}=0\quad \Rightarrow\quad 		\bar{\alpha}_c=-\frac{1}{12n^2}.
	\end{eqnarray} Besides, it is remarkable that $ \bar{\alpha}$ has the upper bound $ \bar{\alpha}=1/12n^2 $, which leads to a simpler and concise  solution, and are displayed in Fig. \ref{phic} and  Fig. \ref{Vphic}:
	\begin{eqnarray}
		&&\bar{\phi}(\bar{y})=\pm 2\sqrt{n}\tanh (\bar{y})\\
		&&\bar{V}(\bar{\phi})=\frac{(5 n+2)}{16 n} \left(\bar{\phi} ^2-4 n\right)^2.
	\end{eqnarray}
	
	 Finally, we also want to explore the distribution of the energy density $\rho=T_{ab}U^aU^b$ along the extra dimension $ y $, where $ U^M=(1/a(y),0,0,0,0,0) $. We introduce the dimensionless energy density:
	\begin{eqnarray}\label{energy}
		\bar{\rho}=\frac{\kappa_6^2}{k_6^2}\rho = \left[(2 n+5 n^2-24 n^3 \alpha -60 n^4 \alpha)\cosh 2\bar{y}+(2 n+5 n^2+24 n^3 \alpha)\right]\text{sech}^4\bar{y}.
	\end{eqnarray}
	\begin{figure}[htbp]\centering
		\begin{minipage}[t]{0.48\textwidth}\centering
			\includegraphics[scale=0.7]{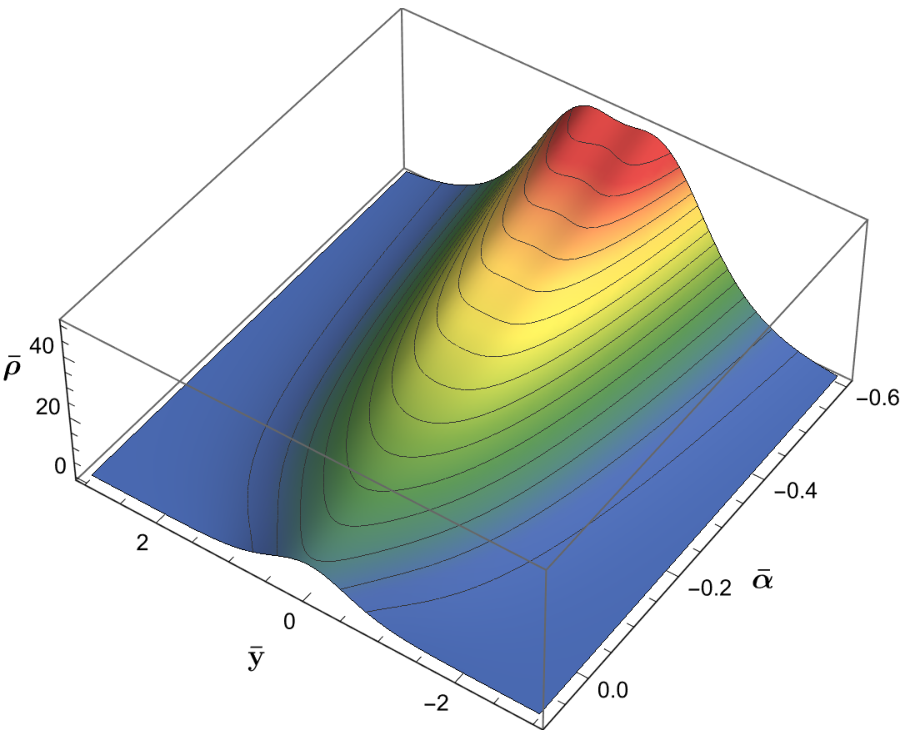}
			\caption{$\bar{\rho}(\bar{y},\bar{\alpha})$ with $ n=1/10$.}\label{rhoyn}
		\end{minipage}
		\begin{minipage}[t]{0.48\textwidth}\centering
			\includegraphics[scale=0.8]{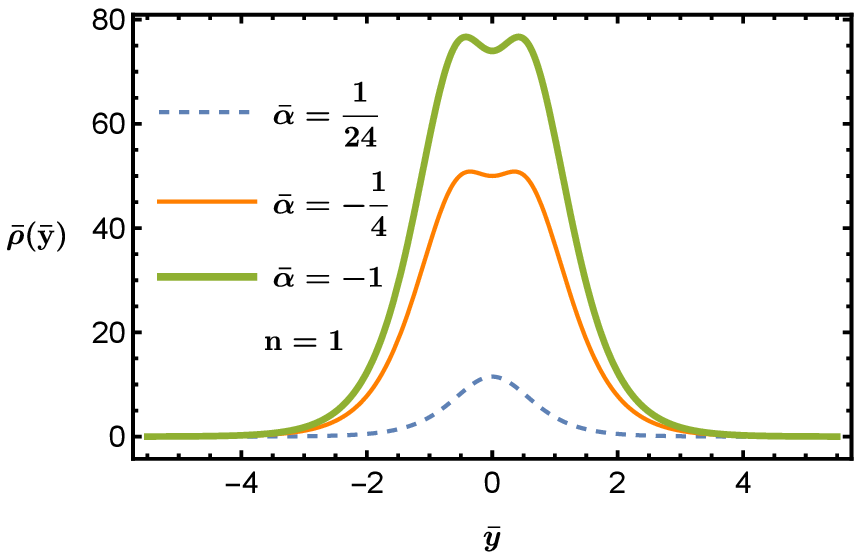}
			\caption{The energy density $ \bar\rho (\bar{y})$ will be split into two different peaks once $ \alpha $ is less than $ \alpha_s $.}\label{split}
		\end{minipage}
	\end{figure} A lot of information can be obtained from the Eq.\eqref{energy}. First, we had better to stress that, the energy density is always positive definite under the condition \eqref{alphacondi}. Second, the energy density behaves differently in various regions of the parameter space of $ n$ and $\bar{\alpha} $. Third, the peak of energy density will increase fastly when $ \alpha<0 $. Importantly, we notice that the energy density of the brane will be split into two different peaks as the value of $ \alpha $ decreases. This behavior is displayed in Fig. \ref{rhoyn} and Fig. \ref{split}. The critical value of $\alpha$ is determined by:
	\begin{eqnarray}%
		\left.\frac{\dd^2\bar{\rho}}{\dd\bar {y}^2}\right|_{\bar{y}=0}=0\quad \Rightarrow\quad\alpha_s=-\frac{2+5n}{24n^2}.
	\end{eqnarray}
	The split of the brane strongly implies that the brane may have some fine inner structure, which may have some physical results. It is worth stressing that, various kinds of reasons will result in a splitting brane, for example, a split brane caused by spacetime torsion is investigated in \cite{yang_thick_2012} and caused by a couple of multiple background scalar fields is study in \cite{bazeia_bloch_2004}.

	\section{Effective Potential}
\begin{figure}[htbp]\centering
		\begin{minipage}[t]{0.45\textwidth}\centering
			\includegraphics[scale=0.8]{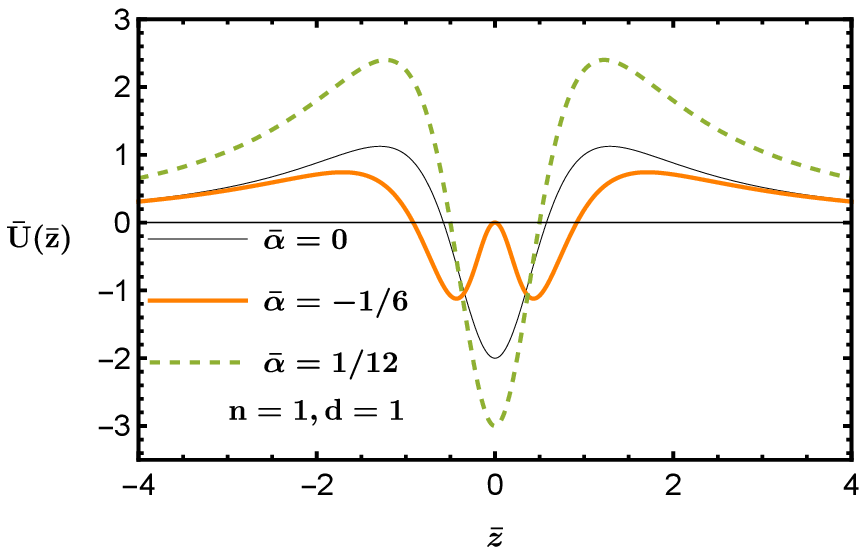}
			\caption{The picture of dimensionless effective potential $ \bar{U{\bar{z}}} =U(z)/k^2, \bar{z}=kz$. A double well caused by Guass-Bonnet  correction term when $ \bar{\alpha}^2 $ is larger.}\label{ualpha1}
		\end{minipage} \hspace{0.05\textwidth}
		\begin{minipage}[t]{0.45\textwidth}\centering
			\includegraphics[scale=0.8]{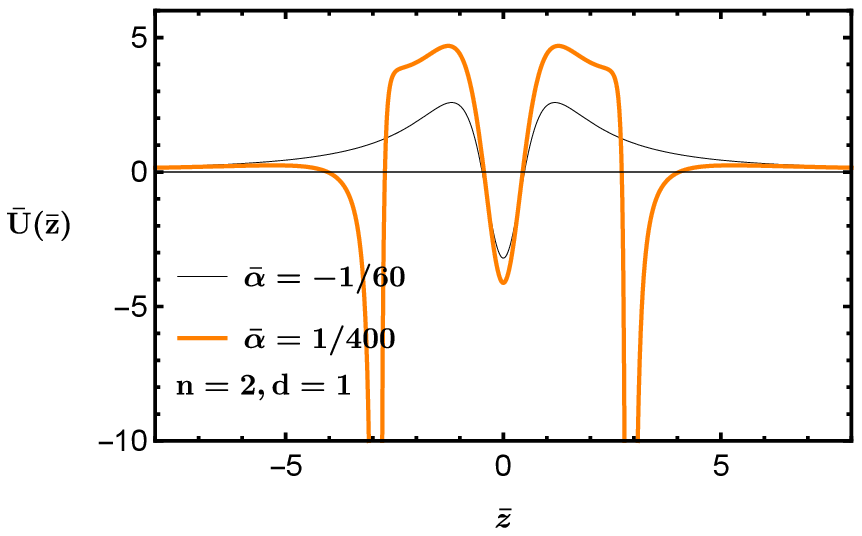}
			\caption{Singularities of the potential may appear for positive $ \bar{\alpha} $ value when $ n>1 $.}\label{ualpha2}
		\end{minipage}
	\end{figure}

	Now we going to discuss the effective potential. With the given warped factor \eqref{ab}, we are able to explore the effective potential \eqref{ep}. The results are depicted in Fig. \ref{ualpha1} and Fig. \ref{ualpha2}. As a comparison, the effective potential of GR is drawn with a thin line in Fig. \ref{ualpha1}, we found that the well goes deeper as $ \bar{\alpha} $ increases. By examining the second derivative with respect to $ \bar{z} $, we found that the potential may split into two wells due to the Gauss-Bonnet correction term as shown in Fig. \ref{ualpha1}. In Fig. \ref{ualpha2}, a strange potential with two singularities is shown, compared to a normal smooth well. Note that although in the above section, we only discuss the brane with $d=1$, the procedure can actually be carried out for general dimensions. Hence, the singularities in D-dimensions (D=4+1+d) are determined by the following equation:
	\begin{eqnarray}
		\xi ^3=2 (1+d) (2+d) n^2 \alpha  (\xi -1) \xi ^{1+n},\qquad\xi=1+z^2\ge1.
	\end{eqnarray} So we can always find two symmetric singularities when $ n>1, \bar{\alpha}>0$.
	
\begin{figure}[htbp]\centering
		\includegraphics[scale=.85]{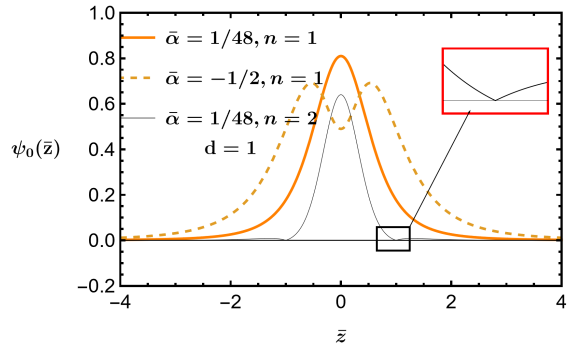}
		\caption{Different possible graviton zero modes. This plot shows the graviton zero modes with one and two peaks, and a special case with non-differentiability at $ \bar{z}=\pm1 $.}\label{psi}
	\end{figure}

	The singularities of the potential should be accused of the non-differentiability of the graviton zero mode $ \psi_0(\bar{z}) $, from which we can see that they can still be normalized. The split effective potential is related to a graviton mode with two different peaks. These behaviors are shown in Fig. \ref{psi}.
	
	\section{Conclusions}
	In summary, we investigate the Gauss-Bonnet thick-brane model in this paper. The behavior of this system under linear tensor perturbation with TT gauge is investigated in detail in this paper. The tensor perturbation equation can be reduced into the Klein-Gordon equation, Helmholz equation, and a second-order linear ordinary differential equation (Schr$\ddot{\mathrm{o}}$dinger-like equation). We conclude that the system is stable in tensor perturbation under certain conditions. Besides, the graviton zero mode is given.

	Furthermore, with a coupling background scalar field and given warp factors, an analytical brane solution in six-dimensional spacetime was constructed. With which we can set a restriction for the Gauss-Bonnet coupling constant in the bulk action. The solution can be simplified a step further. In this case, we claim that the energy density will be split into two peaks with some particular parameter value. The critical value of the coupling constant is given to determine whether it will be split. At the end of the paper, we investigate the effective potential of the Schr$\ddot{\mathrm{o}}$dinger-like equation with respect to various parameters. The potential has a double-well at the bottom which is generated by the Gauss-Bonnet term, compared to one simple well of GR. Moreover, we found that the singularities of the potential should be accused of the non-differentiability of the graviton zero mode $ \psi_0(\bar{z}) $, and the split effective potential is related to a graviton mode with two different peaks.

\begin{acknowledgments}
This work is supported by NSFC with Grants No.11775082.

\end{acknowledgments}

\appendix
	
	\section{Perturbational Quantities}\label{PQ}
	What we want to obtain is the $ (\mu,\nu) $ part of the perturbed field equation\eqref{abspteq}. We choose a method with brute force to evaluate. Theoretically, we can first find out the non-vanishing perturbed Christoffel connection, and then use it to construct the perturbed Riemann curvature tensor. But we do it inversely, we start from the needed quantities like the perturbed Ricci tensor, we might firstly decompose the contraction  $ \delta R_{\mu \nu}=\delta R_{\mu a \nu}{ }^{a} $ into:
	\begin{eqnarray}
		\delta R_{\mu \nu}=\delta R_{\mu \sigma \nu}{ }^{\sigma}+\delta R_{\mu y \nu}{ }^{y}+\delta R_{\mu A \nu}{ }^{A}.
	\end{eqnarray} It is the same for $ H_{\mu\nu }$, but we  need to expand it into more basic quantities firstly:
	\begin{eqnarray}
		\delta H_{\mu\nu}&&=-\delta(4 R_{\mu }{}^{c} R_{\nu c}) + \delta(2 R_{\mu \nu } R) -\delta( 4 R^{cd} R_{\mu c\nu d}) +\delta( 2 R_{\mu }{}^{cde} R_{\nu cde})\notag\\&&-\frac{a^2}{2}\mathcal{L}_{GB}h_{\mu\nu}-\frac{a^2}{2}\eta_{\mu\nu}\delta\mathcal{L}_{GB}.
	\end{eqnarray} And naturally, we may continue to expand the first four terms:
	\begin{eqnarray} 
			\delta(-4 R_{\mu}{}^{c} R_{\nu c})&&=-4(\delta g^{cd}R_{\mu d}R_{\nu c}+g^{cd}\delta R_{\mu d}R_{\nu c}+g^{cd}R_{\mu d}\delta R_{\nu c})\notag\\
			&&=-4(-\frac{h^{\sigma\rho}}{a^2}R_{\mu\rho}R_{\nu\sigma}+g^{\sigma\rho}\delta R_{\mu\rho}R_{\nu\sigma}+g^{\sigma\rho}R_{\mu\rho}\delta R_{\nu\sigma})\notag\\
			&&=4\frac{f^2}{a^2}h_{\mu\nu}-8\frac{f}{a^2}\delta R_{\mu\nu}
	\end{eqnarray}
	\begin{eqnarray} 
			\delta(2R_{\mu\nu}R)=2(R\delta R_{\mu\nu}+R_{\mu\nu}\delta R)
	\end{eqnarray}
	\begin{eqnarray}
			\delta(-4R^{cd} R_{\mu c\nu d})&&=-4(\delta g^{cf}R_{fd} R_{\mu c\nu }{}^d+g^{cf}\delta R_{fd} R_{\mu c\nu }{}^d+g^{cf}R_{fd}\delta  R_{\mu c\nu }{}^d)\notag\\
			=&&-4\left(-\frac{a'^2}{a^2}fh_{\mu\nu}+\frac{a'^2}{a^2}\delta R_{\mu\nu}+\frac{f}{a^2}\delta R_{\mu\sigma\nu}{}^{\sigma}+R_{yy}\delta R_{\mu y\nu}{}^{y}+\frac{s}{b^2}\delta R_{\mu A\nu}{}^{A}\right)
	\end{eqnarray}
	\begin{eqnarray}\label{t4}
			\delta (2R_\mu {}^{cde}R_{\nu cde})&=2(\delta g^{df}R_{\mu cf}{}^eR_{de\nu }{}^c+g^{df}\delta R_{\mu cf}{}^eR_{de\nu }{^c}+g^{df}R_{\mu cf}{}^e\delta R_{de\nu }{}^c)\notag\\
			&=2(\delta g^{\sigma \rho}R_{\mu c\rho }{^e}R_{\sigma e\nu }{^c}+R^f{}_{e\nu }{}^c\delta R_{\mu cf}{^e}+R_{\mu c}{^{de}\delta R_{de\nu }{}^c}).
	\end{eqnarray} The three terms in \eqref{t4} can be expanded as follow:
	\begin{eqnarray}
			\delta g^{\sigma \rho}R_{\mu c\rho }{^e}R_{\sigma e\nu }{^c}=&&-\left(\frac{a'^4}{a^4}+d\left(\frac{a'b'}{ab}\right)^2+\frac{a''^2}{a^2}\right)a^2h_{\mu\nu}\\
			R^f{}_{eb}{}^c\delta R_{acf}{^e}=&& -\frac{a'^2}{a^{2}}\left(\delta R_{\mu \sigma \nu}{ }^{\sigma}-\eta_{\varepsilon \nu} \eta^{\lambda \sigma} \delta R_{\mu \sigma \lambda}{ }^{\varepsilon}\right)-\frac{a''}{a} \delta R_{\mu y \nu}{ }^{y}+a a'' \eta_{\varepsilon \nu}\eta^{yy} \delta R_{\mu 4 4} {}^\varepsilon
			\\&&+\frac{a^2}{b^2} \frac{a' b'}{a b} \eta_{\varepsilon \nu} \eta^{AB}  \delta R_{\mu AB}{}^{\varepsilon}-\frac{a'b'}{a b} \delta R_{\mu E \nu}{}^E\\
			R_{\mu c}{^{de}\delta R_{deb}{}^c}&&=-2 \left(\frac{a'^2}{a^2} \delta R_{\mu \sigma \nu} {}^{\sigma}+ \frac{a^{\prime \prime}}{a} \delta R_{\mu 4 \nu}{ }^{4}+ \frac{a^{\prime} b^{\prime}}{a b} \delta R_{\mu A\nu }{ }^{A}\right).
	\end{eqnarray} Some basic elements, like $ \delta R_{\mu a\nu}{}^b, \eta_{\varepsilon\nu}\eta^{ab}\delta R_{\mu ab}{}^\varepsilon $, appear repeatedly all the way, they can be decomposed into forms like $ \grad_a\delta\Gamma^b{}_{cd} $. After expanding them into coordinate basis, all we need to do is comparing with the  quantities given in \eqref{dchris}.
	\begin{eqnarray}
		&&\grad_\sigma\delta \Gamma^\sigma{}_{\mu\nu}=-\frac12\pd_\sigma\pd^\sigma h_{\mu\nu}+a'a\pd_yh_{\mu\nu}+4\frac{a'}{a}\delta\Gamma^y{}_{\mu\nu}\\
		&&\grad_y\delta \Gamma^y{}_{\mu\nu}=\pd_y\delta\Gamma^y{}_{\mu\nu}-2\frac{a'}{a}\delta\Gamma^y{}_{\mu\nu}\\
		&&\grad_A\delta \Gamma^A{}_{\mu\nu}=\frac12\frac{a^2}{b^2}\pd_A\pd^Ah_{\mu\nu}+d\frac{a'}{a}\delta\Gamma^y{}_{\mu\nu}\\
		&&\grad_\mu\delta \Gamma^\sigma{}_{\sigma\nu}=-\Gamma^a{}_{\mu\sigma}\delta\Gamma^\sigma{}_{a\nu}+\Gamma^\sigma{}_{\mu a}\delta\Gamma^a{}_{\sigma\nu}=-a'^2h_{\mu\nu}\\
		&&\grad_\mu \delta \Gamma^y{}_{y\nu}=a'^2h_{\mu\nu}\\
		&&\grad_\mu \delta \Gamma^A{}_{A\nu}=0.
	\end{eqnarray} Combining the quantities listed above to construct the perturbed Riemann curvature:
	\begin{eqnarray}\label{riemann1}
			&&\delta R_{\mu\sigma\nu}{}^\sigma=-\frac12\pd_\sigma\pd^\sigma h_{\mu\nu}+a'a\pd_yh_{\mu\nu}+4\frac{a'}{a}\delta\Gamma^y{}_{\mu\nu}+a'^2h_{\mu\nu}\\
			&&\delta R_{\mu y \nu}{}^y=\pd_y\delta\Gamma^y{}_{\mu\nu}-2\frac{a'}{a}\delta\Gamma^y{}_{\mu\nu}-a'^2h_{\mu\nu}\\
			&&\delta R_{\mu A\nu}{}^A=\frac12\frac{a^2}{b^2}\pd_A\pd^Ah_{\mu\nu}+d\frac{a'}{a}\delta\Gamma^y{}_{\mu\nu}
	\end{eqnarray}
	And with the same process:
	\begin{eqnarray}
		&&\grad_A\delta \Gamma^\varepsilon{}_{\mu B}=\frac12\pd_A\pd_B h_{\mu}^\varepsilon+\frac12\eta_{A B}b'b\pd_yh_\mu^\varepsilon\\
		&&\grad_y\delta\Gamma^\varepsilon{}_{\mu y}=\frac12\pd_y\pd_yh_\mu^\varepsilon\\
		&&\grad_\sigma\delta\Gamma^\varepsilon{}_{\mu\lambda}=\pd_\sigma\delta\Gamma^\varepsilon{}_{\mu\lambda}+\frac12a'a\eta_{\sigma\mu}\pd_yh_\lambda^\varepsilon+\frac12a'a\eta_{\sigma\lambda}\pd_yh_\mu^\varepsilon+\frac{a'}{a}\delta^\varepsilon_\sigma\delta\Gamma^y{}_{\mu\lambda}\\
		&&\grad_\mu\delta \Gamma^\varepsilon{}_{AB}=0\\
		&&\grad_\mu\delta \Gamma^\varepsilon{}_{yy}=-\frac{a'}{a}\pd_yh_\mu^\varepsilon\\
		&&\grad_\mu\delta \Gamma^\varepsilon{}_{\sigma\lambda}=\pd_\mu\delta\Gamma^\varepsilon{}_{\sigma\lambda}+\frac12a'a\eta_{\sigma\mu}\pd_yh_\lambda^\varepsilon+\frac12a'a\eta_{\mu\lambda}\pd_yh_\sigma^\varepsilon+\frac{a'}{a}\delta^\varepsilon_\mu\delta\Gamma^y{}_{\sigma\lambda}.
	\end{eqnarray}
	We can get another set of perturbed Riemann Curvatue:
	\begin{eqnarray}\label{riemann2}
			&&\eta_{\varepsilon\nu}\eta^{AB}\delta R_{\mu \nu}{}^A=\frac12\pd_A\pd^Ah_{\mu\nu}+\frac12db'b\pd_yh_{\mu\nu}\\
			&&\eta_{\varepsilon\nu}\eta^{44}\delta R_{\mu \nu}{}^y=\frac12\pd_y\pd^yh_{\mu\nu}+\frac{a'}{a}\pd_yh_{\mu\nu}\\
			&&\eta_{\varepsilon\nu}\eta^{\lambda\sigma}\delta R_{\mu\nu}{}^\sigma=\frac12 \pd_\sigma\pd^\sigma h_{\mu\nu}+\frac32a'a\pd_yh_{\mu\nu}+\frac{a'}{a}\delta\Gamma^y{}_{\mu\nu}
	\end{eqnarray}
	With all these elements given above, we can obtain all the necessary perturbed curvature.
	
	In the end, we need to add that:
	\begin{eqnarray}
		\delta R_{\mu y}=\delta R_{\mu A}=\delta R_{yy}=\delta R_{yA}=\delta R_{AB}=0.
	\end{eqnarray} They can be evaluated in a similar way (see Appendix of Ref \cite{cui_tensor_2020}). So the perturbed Ricci scalar vanished as $ \delta R=0 $. And for the perturbed Gauss-Bonnet term:
	\begin{eqnarray}
		\delta \mathcal{L}_{GB}=(-4 R_{a}{}^{c} R_{bc} + 2 R_{ab} R - 4 R^{cd} R_{acbd} + 2 R_{a}{}^{cde} R_{bcde})\delta g^{ab}=0.
	\end{eqnarray} Since the first factor is proportional to $ \delta g_{ab} $ and its derivative, the whole term will finally vanish because of the TT gauge condition \eqref{gauge}.
	
	For the perturbation of energy-momentum tensor $ \delta T_{\mu\nu} $, we only consider the perturbation caused by the fluctuation of the background metric \cite{csaki_universal_2000}. That is to say:
	\begin{eqnarray} 
		\delta T_{\mu\nu}=T_{\mu}^\alpha h_{\alpha\nu}
	\end{eqnarray}
	or practically $ \delta T_{\mu\nu}=1/4\eta^{\alpha\beta}T_{\alpha\beta}h_{\mu\nu} $.


	\bibliographystyle{unsrt}

\end{document}